\documentclass[a4paper]{revtex4}
\usepackage{graphicx}
\usepackage{fancyhdr}
\usepackage{amsmath}
\usepackage{color}
\usepackage{epsfig}
\usepackage{amssymb}
\usepackage{amsmath}
\usepackage{graphicx}

\def\gappeq{\mathrel{\rlap {\raise.5ex\hbox{$>$}}
{\lower.5ex\hbox{$\sim$}}}}
\def\lappeq{\mathrel{\rlap{\raise.5ex\hbox{$<$}}
{\lower.5ex\hbox{$\sim$}}}}

\begin{document}

\title{Imprints of Composite Higgs Models at  $e^+ e^-$ Colliders}

%

\author{D. Barducci$^{a}$, S. De Curtis$^{b}$, S. Moretti$^{c}$, G.M. Pruna$^{d}$} 
\affiliation{
(a) LAPTh,  Universit\`e Savoie Mont Blanc, CNRS, Annecy-le-Vieux, France   \\    (b) INFN and University of Florence, Dept. of Physics and Astronomy, Florence, Italy  \\ (c) School of Physics and Astronomy,  University of Southampton, U.K. \\  (d) Paul Scherrer Institut, Villigen PSI, Switzerland} 
\begin{abstract}
\noindent We test the sensitivity of  a future $e^+e^-$ collider to composite Higgs scenarios encompassing partial compositeness. Besides the detailed study of the Higgs properties, such a machine will have a rich top-quark physics programme mainly in two domains: top property accurate determination at the $t \bar t$ production threshold and search for New Physics  with top quarks above it. In both domains, a composite Higgs scenario can manifest itself via sizable deviations in both cross-section and asymmetry observables.  Herein we discuss such a possibility using a particular realisation, namely the 4-Dimensional Composite Higgs Model. \hspace{3.6cm}  LAPTH-Conf-028/15 

\end{abstract}

\maketitle

\thispagestyle{fancy}


\section{Introduction}

While the  discovery at the Large Hadron Collider (LHC) of a Higgs boson with a mass of 125 GeV is an extraordinary outcome of the CERN machine operations, one of the primary questions that we need to ask now is whether this particle corresponds to the Higgs boson embedded in the Standard Model (SM) or else belongs to some Beyond the SM (BSM) scenario.
In fact, the SM suffers from a key theoretical drawback, the so-called ''hierarchy" problem, pointing out that it could be a low energy effective theory valid only up to some cut-off energy.  The latter can well be at the TeV scale, hence in an energy range accessible at the LHC, so that New Physics (NP) could be discovered at the CERN machine in the coming years.

For this reason many BSM scenarios, which could be revealed at the TeV scale, were proposed in the last decades, the most popular being Supersymmetry, which removes the hiearchy problem through the presence of new states with different spin statistics from the SM ones cancelling (up to residual logarithmic effects) the divergent growth of the SM Higgs mass upon quantum effects. 

Another intriguing possibility is that the Higgs particle may be a composite state arising from some strongly interacting dynamics at a high scale. This will also solve the hierarchy problem owing to compositeness form factors taming such  a divergent behaviour. Furthermore, the now measured light Higgs mass could be  consistent with the fact that such a composite state  arises as a Pseudo Nambu-Goldstone Boson (PNGB) from a particular coset of a global symmetry breaking \cite{Kaplan:1983fs}.
Models with a PNGB  Higgs generally predict modifications of its couplings to both bosons and fermions of the SM, hence the measurement of these quantities represents a powerful way to test the possible non-fundamental nature of the newly discovered state. Moreover, the presence of additional particles in the spectrum of such Composite Higgs Models (CHMs) leads to both mixing effects with SM states as well as new Feynman diagram topologies both of which would represent a further source of deviations from the SM expectations.

In the near future, with the energy increase to 13~TeV for run II, the LHC will be able to test BSM scenarios  more extensively, probing the existence of new particles predicted by BSM extensions to an unprecedented level. Nevertheless the expected bounds, though severe, might  not be conclusive enough to completely exclude such scenarios. As an example,  new gauge bosons predicted by  CHMs with mass larger than $\sim$ 2~TeV could escape the detection of the LHC for several reasons: they could  sizable  interfere with the SM gauge bosons and among themselves, 
so determining a reduction of the number of signal events, or  they could have  sufficiently large widths, due to the opening of extra-fermion decay channels, making their detection critical \cite{DY}.  
Furthermore, concerning the Higgs properties, the LHC[High-Luminosity LHC (HL-LHC)] will not be able to measure the Higgs couplings  to better than   5[2]\% \cite{snowmass} thus leaving room to scenarios, like CHMs, which predict deviations within the foreseen experimental accuracy for natural choices of the compositeness scale $f$ in the TeV range. 

On account of this,  we  will assume  here the case in which LHC will not discover a $W'/Z'$ (or not be able to clearly assess its properties) and also will not discover any extra fermion (or it will discover it with a mass around 700 GeV, which is roughly the present bound, but without any other hint about the theory to which it belongs).  In this situation, an $e^+e^-$ collider could have a great power in  enlightening indirect effects of  BSM physics.

The advantages of  electron-positron colliders with respect to the hadron ones (e.g.,  the cleanliness of the environment, the precision of the measurements, the large number of Higgs bosons produced,  the rich top-quark physics program) make an analysis of its prospects in disentangling the fundamental from the composite nature of the Higgs boson and in testing the partial compositeness of the fermion sector of primary importance.

For these reasons, we will here assess the generic potential of  $e^+e^-$ colliders in testing a particular realisation of CHMs, the so-called 4-Dimensional Composite Higgs Model (4DCHM) of ref.~\cite{4DCHM}. The main Higgs production channels of the 4DCHM were already considered in \cite{LC} for three possible energy stages and different luminosity options of the proposed $e^+e^-$ machines and results were confronted with  the expected experimental accuracies in  various Higgs decay channels.  Here, we want to stress also the potential of  such colliders in discovering  imprints of  partial compositeness in the top-quark sector through  an accurate determination of the top properties at the $t \bar t$ production threshold and in searching for new resonances with top quarks well above their production threshold.  We will consider two processes where a linear collider is expected to crucially complement the LHC physics potential: $e^ +e^ -\to t \bar t H$ and $e^ + e^ -\to t\bar t$, and use the 4DCHM as a calculable prototype describing a PNGB Higgs boson emerging from the $SO(5)\to SO(4)$ breaking.
 
\section{The 4DCHM tested by the $e^+e^-$ collider top-physics program}

A PNGB  Higgs state has non-standard  couplings to both bosons and fermions of the SM, hence a precice measurement of these quantities at present hadronic and future leptonic machines will test the possible non-fundamental nature of the last discovered particle. CHMs generally describe also additional spin-1 and spin-1/2 particles though, both of which  would represent a non-negligible source  of further deviations from the SM expectations.
 
\begin{figure}[h!]
\centering
\includegraphics[width=43mm]{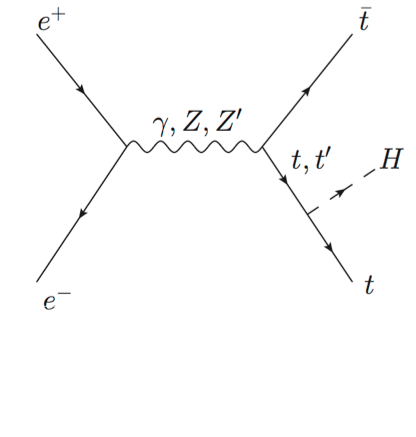}\hspace{0.5cm}
\includegraphics[width=50mm]{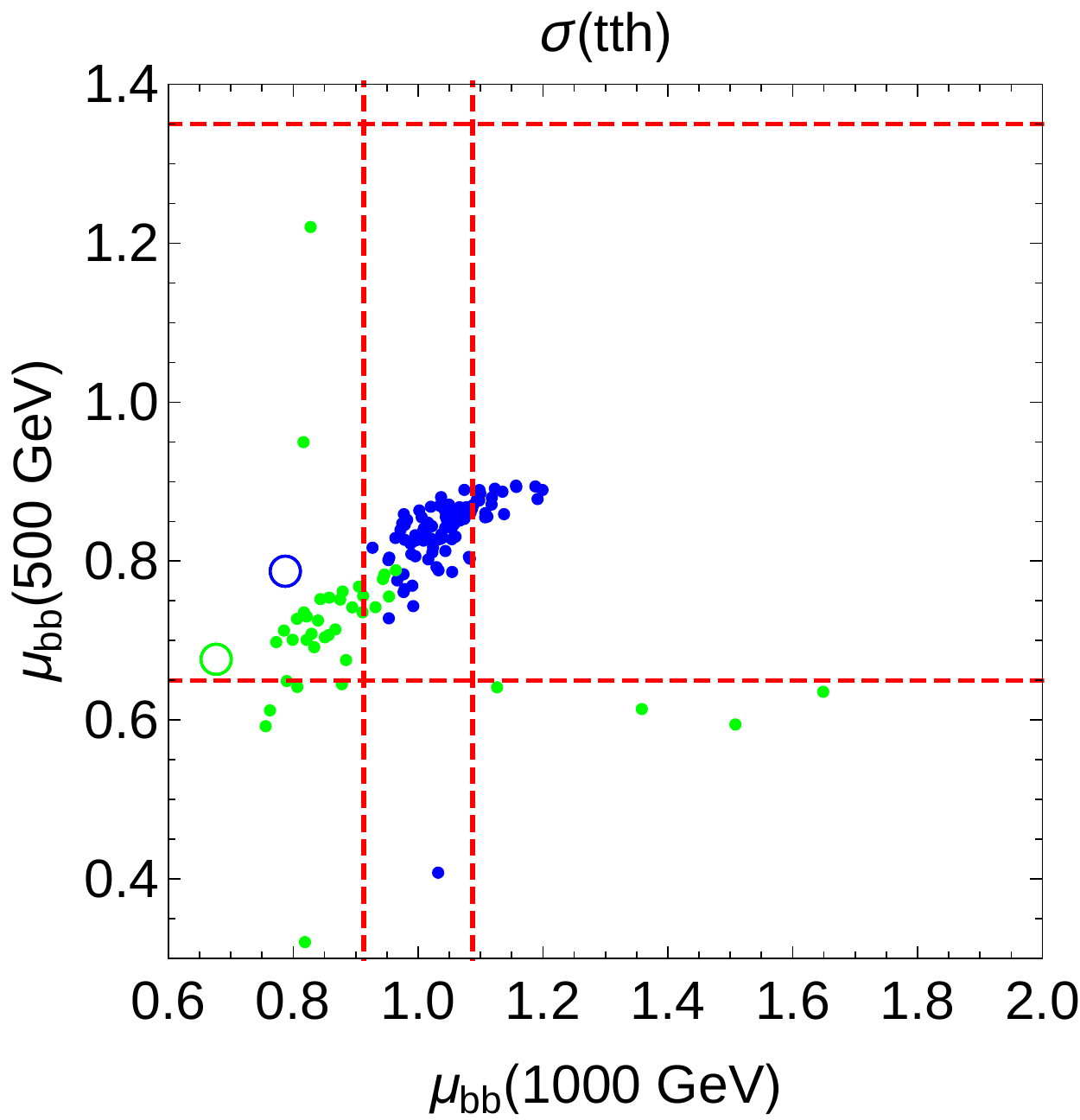}\hspace{0.5cm}
\includegraphics[width=50mm]{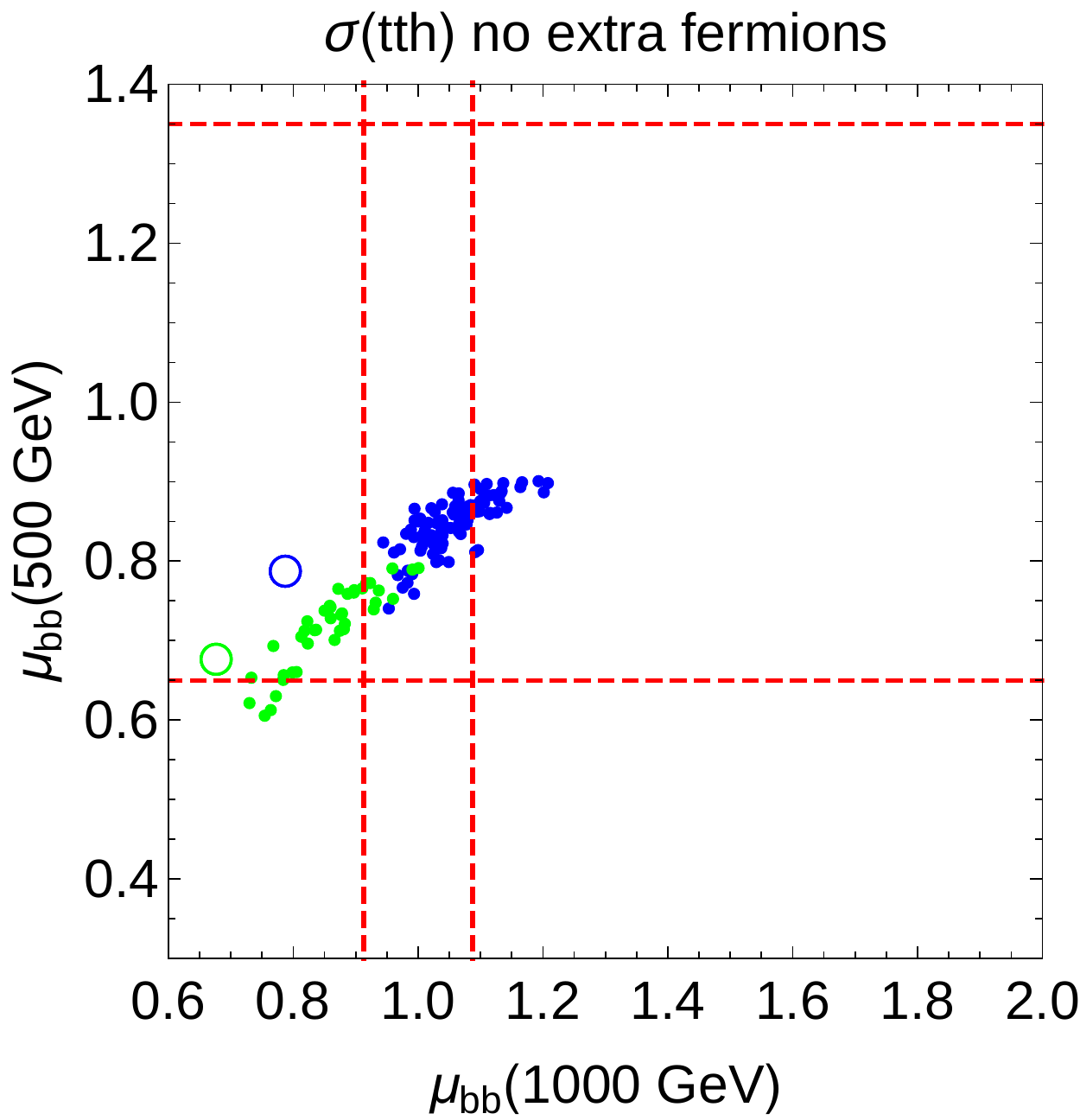}
\caption{Representative Feynman diagram for single Higgs production in association with $t\bar t$ pairs  at an electron-positron collider (left). Correlations among $\mu_{bb}$s evaluated at a future $e^+e^-$ collider for two energy and luminosity stages, in the associated production with $t\bar t$,
with  the inclusion of $t'$ quarks (center) and without these (right). 
Plots are for two 4DCHM benchmarks, with $f=800$ GeV and $g_\rho=2.5$ (green points) and $f=1000$ GeV $g_\rho=2$ (blue points). The red shadowed areas represent the precision limits around the SM expectations according to \cite{peskin}.
The circles represent the values obtained in the "decoupling limit".}
\label{fig:feynHtt}
\end{figure}

Now, focusing onto the 4DCHM \cite{4DCHM}  as an illustrative example representing a description of the minimal additional heavy matter to ensure a finite effective scalar potential,  modifications to the  $e^+e^-\to t \bar t H$ process  (Fig.~\ref{fig:feynHtt} left) arise via the following effects: i) modification of the $Ze^+e^-$ coupling due to the mixing of the $Z$ with the $Z'$s, which however is very small in the parameter range considered; ii) modification of the $Z t \bar t$ coupling which comes from the mixing of the $Z$ with the $Z'$s but also from the mixing  of the top(antitop) with the extra-fermions(antifermions), as expected through  the partial compositeness mechanism; iii)  the exchange of the new particles, namely $Z'$s an $t'$s (both of which can be sizable also because of  interference with the SM states) that, if light enough and/or produced resonantly, could affect the production process significantly.  In 
general CHMs  these extra fermions are predicted to be relatively light, around the TeV scale (so as to avoid a large fine tuning). The present bounds on $t'$ masses depend on their decay branching fractions and are around 700 GeV. The 13~TeV run of the LHC is expected, if no such $t'$ state is discovered, to impose stronger bounds, but, in cases, as the one at hand, of multiple quasi-degenerate new fermion states interfering among themselves and with the same-charge SM quarks, a detailed analysis  is necessary to properly specialize the  bounds extracted from the  experiments. For this reason  we have restricted our parameter scans by only allowing for $t'$  masses  larger  than 700 GeV. Concerning  the $Z'$ mass instead, we will consider configurations corresponding to $M_{Z'}\sim f g_\rho>$ 2~TeV (with $g_\rho$ the $SO(5)$ gauge coupling constant) as requested by the electro-weak  precision tests  and direct searches \cite{DY}.

In the model independent effective approaches, which are generally used for the phenomenological studies of CHMs, the new particle exchange effects are not captured (this corresponds to a "decoupling limit" where the additional particles are made sufficiently heavy). However they can be crucial at both high and moderate center-of-mass energies, $\sqrt{s}$. In \cite{LC} we have shown how quantitative studies of Higgs boson phenomenology in CHMs at future $e^+e^-$ colliders cannot ignore effects from realistic mass spectra. The same conclusion holds true for the extraction of the top Yukawa coupling from $e^+e^-\to t\bar t H$. At the running energy stages of 500 and 1000 GeV this is one of the most important processes to be analysed.  In Fig.~\ref{fig:feynHtt} (center) we plot the correlations among the Higgs signal strengths in the $b\bar b$ decay channel, defined as:
\begin{equation}\label{mu}
\mu_{bb}=\frac{\sigma(e^+e^-\rightarrow H t\bar t)_{\rm 4DCHM}{\rm BR}(H\rightarrow b\bar b )_{\rm 4DCHM}} {\sigma(e^+e^-\rightarrow H t \bar t)_{\rm SM }{\rm BR}(H\rightarrow b \bar b)_{\rm SM}}
\end{equation}
for $\sqrt{s}$ = 500, 1000~GeV. Here we consider  two 4DCHM benchmarks with $f=800$~GeV and $g_\rho=2.5$ (green) and $f=1000$~GeV and $g_\rho=2$ (blue). The points correspond to scans over the fermion sector parameters  as explained in \cite{LC} with the bound $M_{t'}>700$~GeV. The circles in the figure represent the "decoupling limit" obtained by simply rescaling the couplings of the PNGB Higgs to fermions and neglecting the contribution of the extra matter, as done in the aforementioned model independent approaches. In the right panel we plot the results by excluding the $t'$ fermions from the Higgs production process (but keeping the $s$-channel $Z'$ exchange).  Also shown are the expected accuracies in correspondence to  500 and 1000~GeV energies of the collider at  luminosity stages of 500 and 1000 fb$^{-1}$ respectively, of the $b\bar b$ strength as given in \cite{peskin}. As it is clear for these benchmarks, which  correspond to natural choices for the CHM parameters, finite mass effects are sizable and detectable. The biggest deviations of the signal stregths relative to the "decoupling limit" results (mainly at $\sqrt{s}=1$~TeV) are due to the exchange of $t'$ states (from the comparison of the center and right panels) showing that the potential of a future $e^+e^-$ collider in accessing realistic composite Higgs scenarios is very significant also via top-quark processes, with a warning though: the extraction of the top-Yukawa coupling in BSM  schemes, like this one, cannot  simply rely on the rescaling of the coupling predicted by the model.

\begin{figure}[ht]
\centering
\includegraphics[width=55mm]{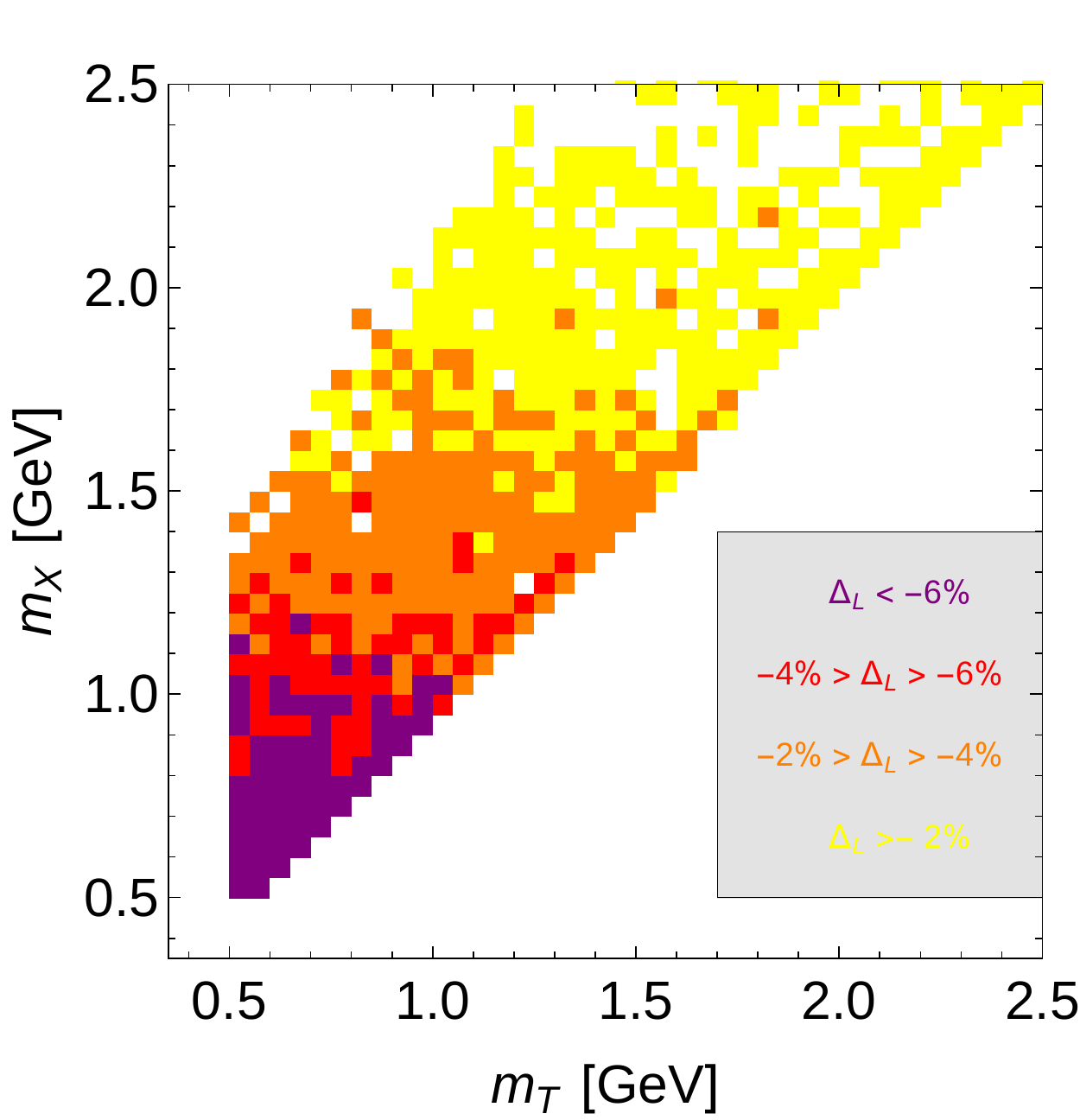}\hspace{1.5cm}
\includegraphics[width=55mm]{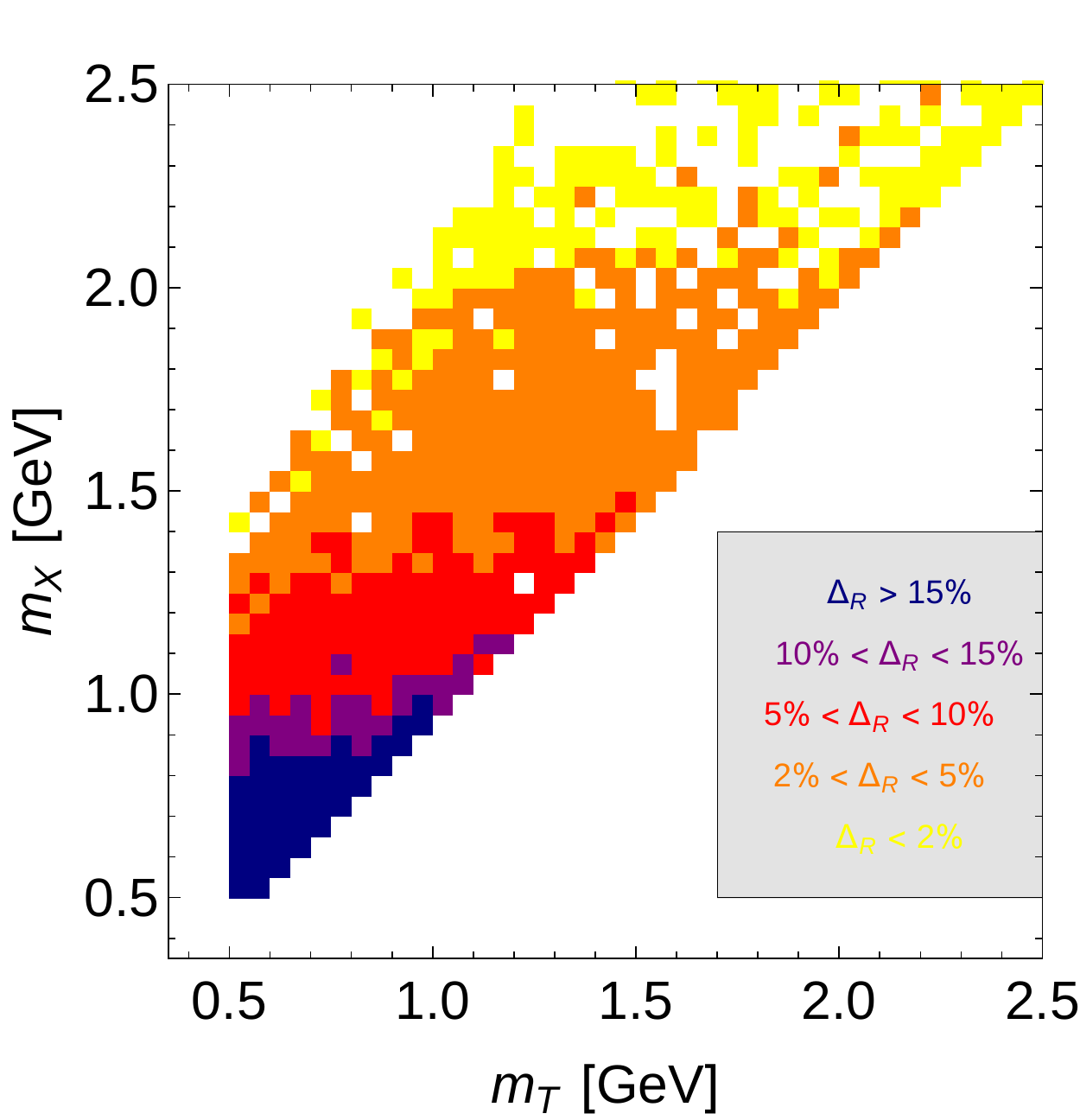}
\caption{
Predicted maximal deviations for $g_L$ (the left-handed coupling of the $Z$ to $t \bar t$) (left) and  $g_R$ (the right-handed one) (right), in the 4DCHM with respect to the SM ones: $\Delta_{L,R}=(g_{L,R}^{4DCHM}-g_{L,R}^{SM})/g_{L,R}^{SM}$.
The regions  correspond to the quoted values of  $\Delta_{L,R}$ from a scan on $0.75<f({\rm TeV})<1.5$, $1.5<g_\rho<3$ (with the constraint $M_{Z'}\sim f g_\rho>2$~TeV) and on the fermion parameters \cite{LC}. Here $T$ and $X$ are the lightest extra-fermions with charge +2/3 and +5/3 respectively.}
\label{gLR}
\end{figure}

The same conclusion holds true for the extraction of the $Z$ couplings to the quark-top by measuring observables related to the $e^+e^-\to t\bar t$ cross-section. In Fig.~\ref{gLR} we separately plot the deviations of $Zt_L\bar t_L$ and $Zt_R\bar t_R$ couplings in the 4DCHM with respect to the SM ones.  The points correspond to a scan in the parameter space of the model and different colours are used for different deviations. We see that even for large masses of the extra fermions $(M_{T,X}>1$~TeV, where $T$ and $X$  are the lightest extra-fermions with charge +2/3 and +5/3, respectively) the modifications can be larger than 5(10)\% for the left(right)-handed coupling.
This result implies a clear and detectable deviation in the $e^+e^-\to t\bar t$ cross-section which is expected to be measured at the per-cent level at a future $e^+e^-$ collider and enforces the importance of this machine in analysing top-physics since these modifications may or may not be detected by the (HL-)LHC measurements, depending on the precision achievable in processes where such  couplings enter (loop corrections to $t\bar t$ production or $Z t\bar t$ events). 

However, as stated, the indirect signals from a composite Higgs scenario are not only encoded in the coupling modifications. Let us then compare the deviations of the 4DCHM (evaluated by taking into account all the ingredients of the model \cite{tt}) with respect to the SM expectations for the following observables: the total cross-section $\sigma(e^+e^-\to t\bar t)$, the forward-backward asymmetry $A_{FB}$ and the spin or top polariation asymmetry  $A_{L}$ (which singles out one final state top, comparing the number of its positive and negative helicities, while summing over the helicities of the antitop).

\begin{figure}[ht]
\begin{center}
\hspace{-0.5cm}
\includegraphics[width=0.33\textwidth]{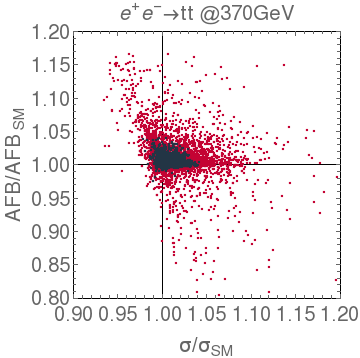}
\includegraphics[width=0.33\textwidth]{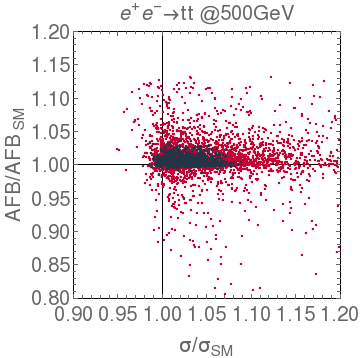}
\includegraphics[width=0.33\textwidth]{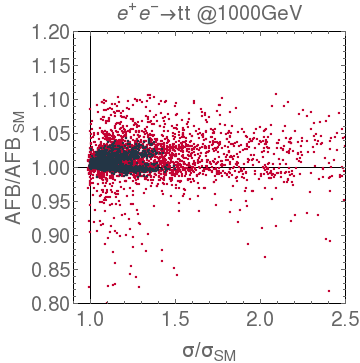}\\\hspace{-0.5cm}
\includegraphics[width=0.33\textwidth]{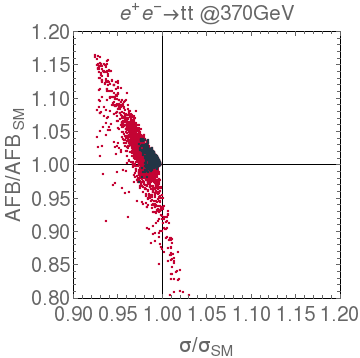}
\includegraphics[width=0.33\textwidth]{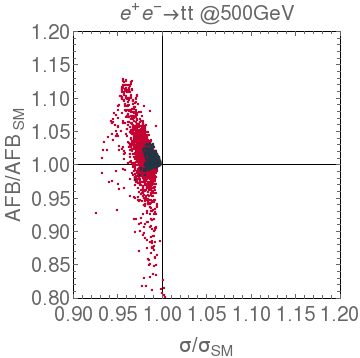}
\includegraphics[width=0.33\textwidth]{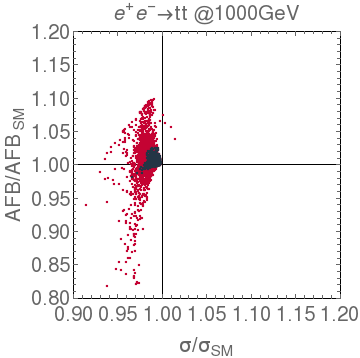}
\vspace{-0.1cm}
\caption{Predicted deviations for the total cross-section and $A_{FB}$, for the process $e^+e^-\to t\bar t$ at 370, 500, 1000~GeV  in the 4DCHM compared with the SM (top panel) and the corresponding ones when the $Z'$s exchange in the $s$-channel is removed (bottom panel). The points correspond to a scan with $f$=0.75-1.5~TeV, $g_\rho$=1.5-3.0 and on the extra-fermion sector parameters as detailed in \cite{tt}. The  black points correspond to the subset with $M_{Z'}>2$~TeV and $M_{t'}>$ 700~GeV.}
\label{xs_AFB}
\end{center}
\end{figure}

We consider here three energy configurations for the $e^+ e^-$ collider: 370, 500, 1000~GeV with unpolarised beams.
The calculation are performed in the Born approximation, Initial State Radiation (ISR) and beamstrahlung are not included (they are not important when considering ratios with respect to the SM expectations).  The most remarkable result of our analysis is the importance of the interference between the SM and the $Z'$s. In Fig.~\ref{xs_AFB}  we show  the correlations between the expected deviations in the cross-section and in $A_{FB}$ for the three energy configurations of the $e^+e^-$ colliders (top panel) and the corresponding ones when the $Z'$s exchange in the $s$-channel is removed (bottom panel).

The effect of the $Z'$ exchange in the cross-section  is always sizable. In fact, already at $\sqrt{s}$ = 370 GeV, the interference tends to compensate the lowering of the cross-section due to the coupling modification (see bottom panels of Fig.~\ref{xs_AFB}) and finally gives a positive contribution which grows with energy, as the interference does, when approaching the mass value of the new resonances (actually there are mainly two nearly-degenerate $Z'$s contributing to the process, namely $Z_2$ and $Z_3$ \cite{DY}). Deviations up to 50\% are expected in the total cross-section at $\sqrt{s}$ = 1000 GeV while the effect on the $A_{FB}$ is less evident.  In fact, for this observable we are dividing by the total cross-section and this washes out the large $Z'$ interference dependence. Anyway, we expect that such deviations are all detectable (already at 370 GeV) beyond  the experimental errors, which are generally claimed to be at the level of  percent or even smaller for the cross-section \cite{AAJ}. A hard task is to disentangle the different effects, in order to extract information on the type of BSM physics producing such deviations.

\begin{figure}[ht]
\begin{center}
\hspace{-0.5cm}
\includegraphics[width=0.33\textwidth]{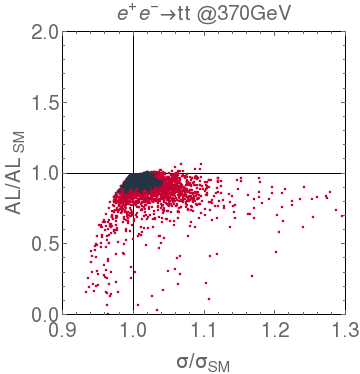}
\includegraphics[width=0.33\textwidth]{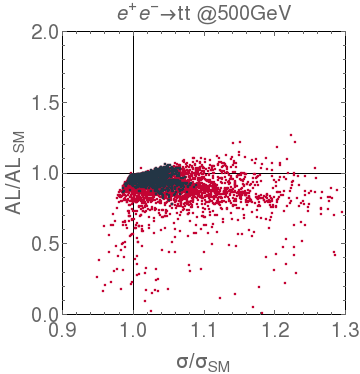}
\includegraphics[width=0.33\textwidth]{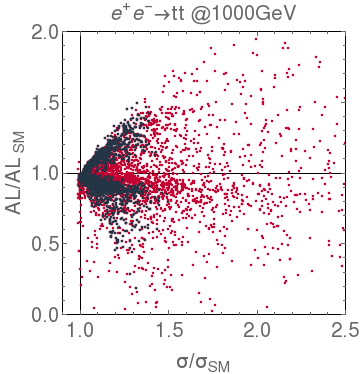}
\vspace{-0.1cm}
\caption{Predicted deviations of the cross-section and  $A_{L}$ for the process $e^+e^-\to t\bar t$ at 370, 500, 1000 GeV  in the 4DCHM compared with the SM. The points correspond to $f$=0.75-1.5~TeV, $g_\rho$=1.5-3.0. The color code is the same of Fig.~\ref{xs_AFB}. }
\label{xs_AL}
\end{center}
\end{figure}

Among the  asymmetries, $A_L$ deserves particular attention. In fact,  it is sensitive to the relative sign of the left- and right-handed couplings of the  $Z$ and $Z'$s to the top pair. This property was already pointed out in \cite{ken} where  the production of $t \bar t$ pairs at the LHC within the 4DCHM was studied. As shown therein, $A_L$ is unique in offering the chance to separate $Z_2$ and $Z_3$ as the two spin-1 states contribute to this asymmetry in opposite directions. In Fig.~\ref{xs_AL}  the expected deviations are shown. Notice that, for $\sqrt{s}=1000$ GeV, where the interference of $Z_2$ and $Z_3$ with $\gamma,Z$ is largest, the two contributions appear visible in opposite directions and the deviations can reach 50\%.
\vspace{-0.3cm}

In summary, under the described circumstances, wherein a detection of new $Z'$s or $t'$s from a CHM  cannot either be established with enough significance at the LHC or else this machine cannot resolve nearby resonances (as typically predicted  by CHMs  like the one considered here), future leptonic colliders also afford one with the ability to combine charge and spin
 asymmetries together with the total cross-section for the process $e^+e^-\to t \bar t$, thereby enabling one 
to increase significances, to the extent of possibly claiming an indirect discovery of a CHM structure of EW Symmetry Breaking (EWSB), even for
$Z'$ masses well beyond the kinematic reach of the leptonic accelerator.   Finally, also the process $e^+e^-\to H t \bar t$ could indirectly discover the imprint of extra fermions (if still undetected) and of top partial compositeness.
\vspace{-0.2cm}
\begin{acknowledgments}
\noindent SM is financially supported in part by the NExT 
Institute and JSPS 
in the form of a Short Term Fellowship for Research in Japan
(Grant Number S14026). 
\end{acknowledgments}

\end{document}